\documentclass[9pt]{article}
\usepackage{spconf,amsmath,graphicx,hyperref}

\usepackage{amsmath}
\usepackage[most]{tcolorbox}
\usepackage{xcolor}

\definecolor{royalblue}{rgb}{0.21,0.49,0.74}
\newcommand{\makeblue}[1]{\textcolor{royalblue}{#1}}
\definecolor{lightpurple}{HTML}{9999FF}
\usepackage{multirow}
\usepackage[table]{xcolor}
\usepackage{amssymb}
\usepackage{pifont}
\usepackage{makecell}
\usepackage{booktabs}
\usepackage{makecell}
\usepackage{color}

\usepackage{cite}
\usepackage{graphicx}    
\usepackage{subcaption}  
\usepackage{hyperref}
\usepackage{booktabs}
\usepackage{graphicx}
\usepackage{makecell}
\usepackage{multirow}
\usepackage{caption}
\usepackage{bbm}
\usepackage{float}
\usepackage{orcidlink}

\makeatletter
\g@addto@macro\@maketitle{%
  \vspace{-1.0em}%
  \begin{minipage}{\textwidth}
    \centering
    \includegraphics[width=0.78\textwidth]{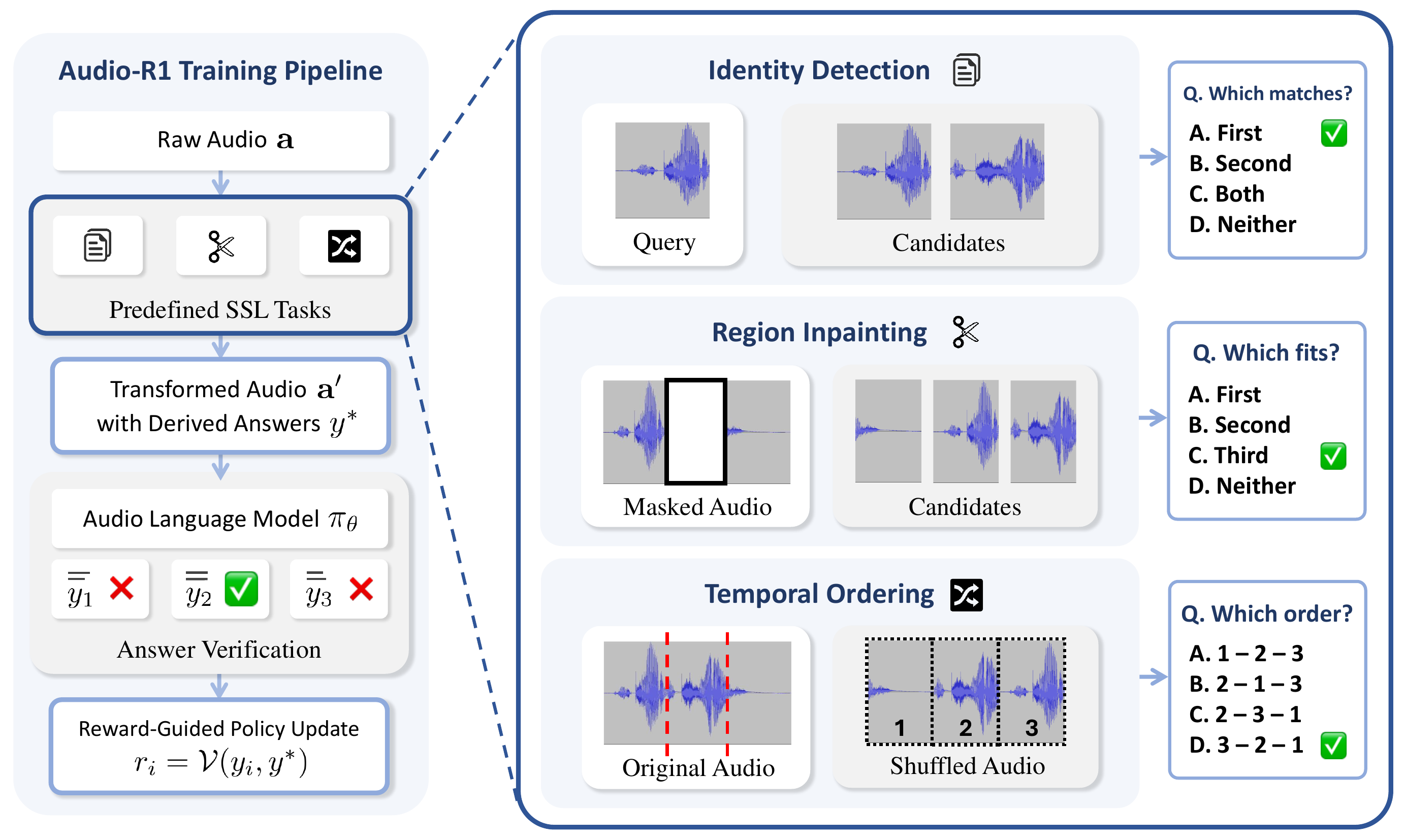}
    \vspace{-0.6em}
    \captionof{figure}{
      \textbf{Audio-R1 framework and self-supervised task construction.}
      Left: Raw audio is transformed into task instances with answers known by construction, which are used to verify model responses and compute rewards for policy optimization. Right: Examples of three different Audio-R1 tasks (Identity Detection, Region Inpainting, and Temporal Ordering).
    }
    \label{fig:audio_r1_overview}
  \end{minipage}
  \vspace{1.0em}%
}
\makeatother

\title{Audio-R1: Audio Reasoning through Reinforcement Learning on Self-Supervised Tasks}
\name{Jihoon Jeong\,\textsuperscript{1,2}, Mirco Ravanelli\,\textsuperscript{1,3}, Zhepei Wang\,\textsuperscript{4}, Cem Subakan\,\textsuperscript{1,2,3}}
\address{\textsuperscript{1}Mila-Qu\'{e}bec AI Institute \quad \textsuperscript{2}Universit\'{e} Laval \quad \textsuperscript{3}Concordia University \quad \textsuperscript{4}Adobe Research}
\begin{document}
\ninept
\maketitle

\begin{abstract}
Reasoning is increasingly important for large audio-language models (LALMs), yet remains relatively underexplored compared with text-only and vision-language models.
We introduce Audio-R1, a reinforcement learning (RL) post-training framework that revisits audio self-supervised learning as a source of verifiable rewards for improving audio understanding and reasoning task performance.
Audio-R1 uses answers automatically determined during the construction of self-supervised tasks to verify model responses and compute rewards.
Experiments across multiple LALMs and diverse audio benchmarks show consistent performance gains, while ablations demonstrate the benefits of each self-supervised task and additional gains from joint training.
While both supervised fine-tuning and Audio-R1 improve self-supervised task accuracy, Audio-R1 further improves performance on downstream tasks distinct from the training tasks.
These results highlight the potential of RL post-training with self-supervised audio tasks to improve downstream performance and reasoning-trace quality.
\end{abstract}


\begin{tcolorbox}[
    colback=gray!8,
    colframe=gray!50,
    arc=2mm,
    boxrule=0.5pt,
    boxsep=0.8mm,
    top=0.5mm,
    bottom=0.5mm
]
\small
\centering
Demos and examples are available on our
\href{https://ijihoon98.github.io/Audio-R1/}{\textcolor{royalblue}{\textbf{project page}}}.
\end{tcolorbox}

\section{Introduction}
\label{sec:intro}

\vspace{-2mm}

Large audio-language models (LALMs) have shifted audio understanding toward a unified modeling paradigm, enabling a single model to support diverse tasks, from automatic speech recognition to spoken interaction \cite{tang2023salmonn, xu2025qwen25omni}.
LALMs can also solve reasoning tasks with audio inputs by leveraging the reasoning capabilities of their underlying language models \cite{ma2025audio, goel2025audio, xu2025qwen3}.
However, their audio reasoning capabilities remain underdeveloped relative to those of text-only and vision-language models, with recent studies showing that explicit reasoning does not consistently improve performance and can even underperform direct answering \cite{ma2025audio, li2025reinforcement, fan2026incentivizing}.
We therefore investigate how to improve both audio understanding and reasoning in LALMs.

Supervised fine-tuning (SFT) and reinforcement learning (RL) are central approaches to improving reasoning performance \cite{zhifei2025audio, li2025reinforcement, Bonzi2026_reasoning}, but applying them to LALMs presents two key challenges.
First, collecting high-quality supervised reasoning data for SFT is costly and hard to scale across diverse, complex real-world audio tasks \cite{he2026audio}.
Second, answer-correctness rewards on conventional audio question-answering tasks do not ensure that models rely on acoustic evidence, motivating automatically constructed tasks whose solutions depend on the content and structure of the audio \cite{tian2025step, jeong2026listen}.
These limitations motivate a scalable post-training approach that derives supervision directly from the content and structure of audio.

Building on this idea, we propose Audio-R1, an RL post-training framework that derives rewards from self-supervised audio tasks, as shown in Fig.~\ref{fig:audio_r1_overview}.
Inspired by audio self-supervised learning (SSL) \cite{saeed2021contrastive, huang2022masked, carr2021self, ravanelli2020multi}, we construct tasks with automatically determined answers and use them to verify LALM responses, turning answer correctness into rewards for policy optimization.
Specifically, we design three audio tasks, \textit{Identity Detection}, \textit{Region Inpainting}, and \textit{Temporal Ordering}, that cover audio information at different granularities, from individual audio samples to longer time spans.

Experiments across diverse audio benchmarks show that Audio-R1 consistently enhances LALM capabilities. 
Multi-task RL training consistently outperforms all single-task variants on all three self-supervised tasks and on average across the three downstream benchmarks, suggesting complementary learning signals across tasks.
Although SFT and Audio-R1 both improve accuracy on the given tasks, SFT provides only modest downstream transfer, whereas Audio-R1 achieves substantially larger downstream improvements.
An audio-conditioned judge also assigns Audio-R1 higher reasoning-trace quality scores than the base model and SFT, highlighting the importance of the learning approach for both generalization and reasoning quality.



Our contributions can be summarized as follows:
\begin{itemize}

    \item We introduce Audio-R1, an RL post-training framework that uses three self-supervised audio tasks as a source of verifiable rewards to improve LALM performance (Sec.~\ref{sec:methodology}).


    \item We demonstrate that, when Audio-R1 is applied to four different base models, consistent gains are observed across three benchmarks (MMAR, MMAU-mini, and MMAU-Pro). Task ablations show gains from each SSL task individually, with further improvements from joint training (Sec.~\ref{subsec:rq1} and~\ref{subsec:rq2}).
    

    \item We provide evidence that Audio-R1 improves reasoning-trace quality relative to the base model through an LLM-as-judge evaluation (Sec.~\ref{subsec:rq3}).
    

\end{itemize}

\section{Related Work}
\label{sec:related_work}

\noindent \textbf{Self-Supervised Learning.}
SSL has been widely used to learn general audio representations from unlabeled recordings through objectives such as contrastive learning, masked prediction, self-distillation, and acoustic-token prediction \cite{saeed2021contrastive, ravanelli2020multi, baevski2020wav2vec, hsu2021hubert, chen2022wavlm, niizumi2021byola, huang2022masked}.
By deriving training targets from the recordings themselves, these approaches enable learning from large-scale audio data without manual annotation.
While prior work focuses on pre-training, the use of SSL for post-training of LALMs remains underexplored \cite{wang2025jigsaw, xie2026sslr1}.
Audio-R1 extends SSL to this setting by using task answers derived from audio to compute rewards, \textit{shifting the focus from learning representations to improving audio reasoning beyond the training tasks.}

\medskip
\noindent \textbf{Reinforcement Learning-Based Reasoning for Large Audio-Language Models.}
Recent studies have applied RL to LALMs to improve audio understanding and reasoning \cite{li2025reinforcement, rouditchenko2025omni}.
Approaches include answer-based rewards from audio question-answering datasets and process-level feedback that encourages structured, consistent, or acoustically grounded reasoning \cite{wen2025sari, wu2026audio, tian2025step, jeong2026listen}.
However, their reliance on human-annotated question-answer pairs, model-generated reasoning traces, or task-specific evaluators can limit scalability and task diversity.
In contrast, Audio-R1 derives its supervision \textit{from self-supervised task construction rather than externally supplied answers or judgments}, using the resulting task answers to compute RL rewards.

\section{Methodology}
\label{sec:methodology}

Audio-R1 is a generic RL post-training framework that derives verifiable rewards from self-supervised audio tasks (Fig.~\ref{fig:audio_r1_overview}).
These three tasks encourage the use of acoustic, temporal, and semantic cues.
In this section, we first introduce related RL preliminaries in Sec. \ref{subsec:preliminary}. 
We then detail our pipeline, including task and reward designs, in Sec. \ref{subsec:task_design} and Sec. \ref{subsec:reward_design}.

\subsection{Preliminary}
\label{subsec:preliminary}

\noindent
\textbf{Reinforcement Learning with Verifiable Rewards.}
Recent work on reasoning-oriented post-training has explored reinforcement learning with verifiable rewards (RLVR) \cite{lambert2024tulu}, which evaluates model responses using rule-based verifiers rather than learned reward models.
Unlike SFT, which imitates predefined demonstrations, RLVR encourages exploration through outcome-based feedback.
In Audio-R1, we use Group Relative Policy Optimization (GRPO) \cite{shao2024deepseekmath} to optimize the policy under this framework.
For each prompt $x$, we sample $L$ responses $\{y_i\}_{i=1}^{L}$ from the policy $\pi_{\theta}(\cdot\mid x)$ and compute their rewards as
\begin{equation}
    r_i = \mathcal{V}(y_i, y^{*}), \qquad i=1,\ldots,L.
\end{equation}
where $\mathcal{V}$ is a rule-based reward function and $y^{*}$ is the ground-truth answer.
GRPO computes each response advantage as $\hat{A}_i=(r_i-\mu_r)/(\sigma_r+\epsilon)$, where $\mu_r$ and $\sigma_r$ are the mean and standard deviation of the group rewards, and $\epsilon$ ensures numerical stability.
These advantages guide policy updates to favor higher-reward responses, enabling learning without supervision of the reasoning process.


\subsection{Task Design}
\label{subsec:task_design}


As illustrated in Figure~\ref{fig:audio_r1_overview}, we construct three self-supervised audio tasks whose answers are known by construction, enabling scalable data generation and verification of deterministic responses without human annotations or external supervision.
Although simple to construct, the tasks are designed to probe complementary aspects of audio understanding: cross-clip correspondence, local contextual compatibility, and temporal structure.
For each task, we denote the audio input by $\mathbf{a}^{\prime}$, which may consist of multiple clips or segments, and the target answer by $y^{*}$, with a task-specific subscript.
Depending on the task, $y^{*}$ takes the form of an index set, a single index, or an ordered sequence of indices.

\smallskip
\noindent
\textbf{Identity Detection.}
The input $\mathbf{a}^{\prime}=(\mathbf{a},\mathcal{C}_{\mathrm{id}})$ consists of an anchor waveform $\mathbf{a}\in\mathbb{R}^{T}$ and candidate clips $\mathcal{C}_{\mathrm{id}}=\{\mathbf{c}_i\}_{i=1}^{N_{\mathrm{id}}}$ drawn from the same or different source recordings.
Let $\rho(\cdot)$ denote the source-recording identity. The model identifies the subset of candidates from the same recording as the anchor:
\begin{equation}
    y^{*}_{\mathrm{id}}
    =
    \left\{
        i \in \{1,\ldots,N_{\mathrm{id}}\}
        \;\middle|\;
        \rho(\mathbf{c}_i)=\rho(\mathbf{a})
    \right\}.
\end{equation}
In our implementation, each instance contains an anchor and two candidates ($N_{\mathrm{id}}=2$), all 5 seconds long.
Positives are exact copies of the anchor crop; negatives are sampled uniformly without replacement from the same pool, excluding the anchor.

\smallskip
\noindent
\textbf{Region Inpainting.}
Given an audio waveform $\mathbf{a}\in\mathbb{R}^{T}$, we remove a contiguous segment $\mathbf{a}_{[t_s:t_e]}$ and retain the surrounding context.
This context and candidate segments $\mathcal{C}_{\mathrm{inp}}=\{\mathbf{c}_i\}_{i=1}^{N_{\mathrm{inp}}}$ form the input $\mathbf{a}^{\prime}$.
The model identifies the candidate that originally occupied the missing region:
\begin{equation}
    y^{*}_{\mathrm{inp}} = \arg\max_{i\in\{1,\ldots,N_{\mathrm{inp}}\}}
    \mathbbm{1}\!\left[
        \mathbf{c}_i=\mathbf{a}_{[t_s:t_e]}
    \right].
\end{equation}
In our implementation, we replace the middle segment with silence.
The masked clip is presented alongside four candidates ($N_{\mathrm{inp}}=4$): the removed segment and three distractors of the same length, center-cropped from other recordings.

\smallskip
\noindent
\textbf{Temporal Ordering.}
We divide a continuous audio recording into an ordered sequence of $N_{\mathrm{ord}}$ segments $\mathcal{C}_{\mathrm{ord}}=(\mathbf{c}_1,\ldots,\mathbf{c}_{N_{\mathrm{ord}}})$.
A random permutation $\boldsymbol{\pi}$ produces the input $\mathbf{a}^{\prime}=(\mathbf{c}_{\pi_1},\ldots,\mathbf{c}_{\pi_{N_{\mathrm{ord}}}})$.
The model recovers the original chronological order by predicting
\begin{equation}
    y^{*}_{\mathrm{ord}}=\operatorname{argsort}(\boldsymbol{\pi}),
\end{equation}
which lists the presented segment indices in chronological order.
In our implementation, three segments ($N_{\mathrm{ord}}=3$) are presented as separate audio clips.
Answer options are randomly sampled from the six possible permutations, ensuring inclusion of the correct order.
The prompts and judge rubric are provided in Appendix~\ref{app:instructions}.

\subsection{Reward Design}
\label{subsec:reward_design}

Building on the tasks designed to encourage the use of audio evidence, we use final-answer correctness and format compliance as learning signals for reasoning traces.
For each generated response $y_i$, the rule-based reward function $\mathcal{V}$ combines these two components:
\begin{equation}
    \mathcal{V}(y_i,y^{*})
    = R_{\mathrm{acc},i} + \lambda R_{\mathrm{format},i},
\end{equation}
where $\lambda$ controls the contribution of format compliance relative to answer correctness.

\smallskip
\noindent
\textbf{Accuracy Reward.}
We compute the accuracy reward by comparing the model's answer with the answer derived from task construction.  
Since each task is formulated as a multiple-choice question, we assign the binary reward $R_{\mathrm{acc},i}=\mathbbm{1}[\hat{y}_i=y^{*}]$, where $\hat{y}_i$ is the final answer extracted from response $y_i$, and $y^{*}$ is the ground-truth answer defined in Sec.~\ref{subsec:task_design}.

\smallskip
\noindent
\textbf{Format Reward.}
The format reward $R_{\mathrm{format},i}$ is a binary indicator of whether the response follows the required structure.
Specifically, the reasoning must be enclosed within \texttt{<think>} and \texttt{</think>}, followed by the final answer enclosed within \texttt{<answer>} and \texttt{</answer>}.
We assign $R_{\mathrm{format},i}=1$ when response $y_i$ follows this format and $R_{\mathrm{format},i}=0$ otherwise.

\section{Experimental Setup}
\label{sec:experiments}

\noindent
\textbf{Dataset Settings.}
We construct a diverse audio pool of approximately 24K source recordings from MELD~\cite{poria2019meld}, LibriSpeech~\cite{panayotov2015librispeech}, VoxCeleb2~\cite{chung2018voxceleb2}, Multispeaker~\cite{zhifei2025audio}, YouTube-8M~\cite{abu2016youtube}, and CoVoST~2~\cite{wang2020covost}.
We convert all audio to 16~kHz mono and use it to construct training instances for the three self-supervised tasks described in Sec.~\ref{subsec:task_design}.

\smallskip
\noindent
\textbf{Training Setup.}
We apply LoRA to the LLM backbone and audio aligner while keeping the encoder frozen, and optimize with GRPO using KL regularization against the reference policy ($\beta=0.04$).
We set the format reward weight to $\lambda=0.2$ (Sec.~\ref{subsec:reward_design}).
For each prompt, we sample eight responses at a temperature of $0.8$ with top-$p=0.9$.
We train for one epoch on 8$\times$H200 GPUs with a learning rate of $5\times10^{-6}$, 3\% linear warmup, and no weight decay.

\smallskip
\noindent
\textbf{Benchmarks.}
We evaluate transfer to established audio perception and reasoning benchmarks: MMAR~\cite{ma2025mmar}, MMAU-mini~\cite{sakshi2024mmau}, and MMAU-Pro~\cite{kumar2026mmaupro}.
We additionally construct held-out test sets for the three self-supervised tasks using MMAU-mini and MMAR recordings to evaluate the capabilities directly targeted during post-training.


\begin{table}[H]
\centering
\caption{
\textbf{Accuracy comparison on MMAR, MMAU-mini, and MMAU-Pro.}
The top, middle, and bottom sections report closed-source models, RL-finetuned models, and open-weight backbones. 
In the bottom section, each base model is followed by a “+ Audio-R1” row showing its performance after applying Audio-R1.
}
\vspace{-2mm}
\label{tab:main_benchmark_comparison}

\small
\setlength{\tabcolsep}{5.0pt}
\renewcommand{\arraystretch}{1.10}

\resizebox{\columnwidth}{!}{
\begin{tabular}{lccc}
\toprule
\textbf{Model}
& \textbf{MMAR ($\uparrow$)}
& \textbf{MMAU-mini ($\uparrow$)}
& \textbf{MMAU-Pro ($\uparrow$)} \\
\midrule

\rowcolor{gray!12}
\multicolumn{4}{l}{\textit{Closed-source Models}} \\

GPT-4o mini Audio~\cite{hurst2024gpt}
& 51.0 & 59.6 & 48.3 \\

GPT-4o Audio~\cite{hurst2024gpt}
& 62.8 & 62.5 & 52.5 \\

Gemini 2.0 Flash~\cite{google2025gemini20flash}
& 65.5 & 70.5 & 55.7 \\

Gemini 2.5 Flash~\cite{comanici2025gemini}
& 68.9 & 71.8 & 59.2 \\


\midrule
\rowcolor{gray!12}
\multicolumn{4}{l}{\textit{Models Finetuned with RL}} \\

Audio-Reasoner~\cite{zhifei2025audio}
& 37.5 & 67.7 & 39.5 \\

Omni-R1~\cite{rouditchenko2025omni}
& 63.4 & 76.0 & -- \\

Audio-Thinker~\cite{wu2026audio}
& 65.3 & 76.5 & -- \\

\midrule
\rowcolor{gray!12}
\multicolumn{4}{l}{\textit{Open-weight Models w/ and w/o Audio-R1}} \\



\addlinespace[2pt]
Gemma4-e4b~\cite{gemmateam2026gemma4}
& 50.0 & 62.3 & 48.7 \\

\rowcolor{gray!6}
\quad \textbf{+ Audio-R1}
& \textbf{52.8} {\scriptsize(\makeblue{+2.8})}
& \textbf{64.1} {\scriptsize(\makeblue{+1.8})}
& \textbf{50.5} {\scriptsize(\makeblue{+1.8})} \\

\addlinespace[2pt]
AF-Next-Think~\cite{ghosh2026audioflamingonext}
& 56.0 & 74.5 & 54.1 \\

\rowcolor{gray!6}
\quad \textbf{+ Audio-R1}
& \textbf{57.0} {\scriptsize(\makeblue{+1.0})}
& \textbf{77.2} {\scriptsize(\makeblue{+2.7})}
& \textbf{56.0} {\scriptsize(\makeblue{+1.9})} \\

\addlinespace[2pt]
Qwen2.5-Omni~\cite{xu2025qwen25omni}
& 58.2 & 73.4 & 55.1 \\

\rowcolor{gray!6}
\quad \textbf{+ Audio-R1}
& \textbf{62.6} {\scriptsize(\makeblue{+4.4})}
& \textbf{77.2} {\scriptsize(\makeblue{+3.8})}
& \textbf{58.7} {\scriptsize(\makeblue{+3.6})} \\

\addlinespace[2pt]
Qwen3-Omni~\cite{xu2025qwen3}
& 68.5 & 74.9 & 61.5 \\

\rowcolor{gray!6}
\quad \textbf{+ Audio-R1}
& \textbf{70.3} {\scriptsize(\makeblue{+1.8})}
& \textbf{77.5} {\scriptsize(\makeblue{+2.6})}
& \textbf{63.6} {\scriptsize(\makeblue{+2.1})} \\



\bottomrule
\end{tabular}
}
\vspace{-4mm}
\end{table}

\begin{figure}[H]
    \centering
    \includegraphics[width=0.8\linewidth]{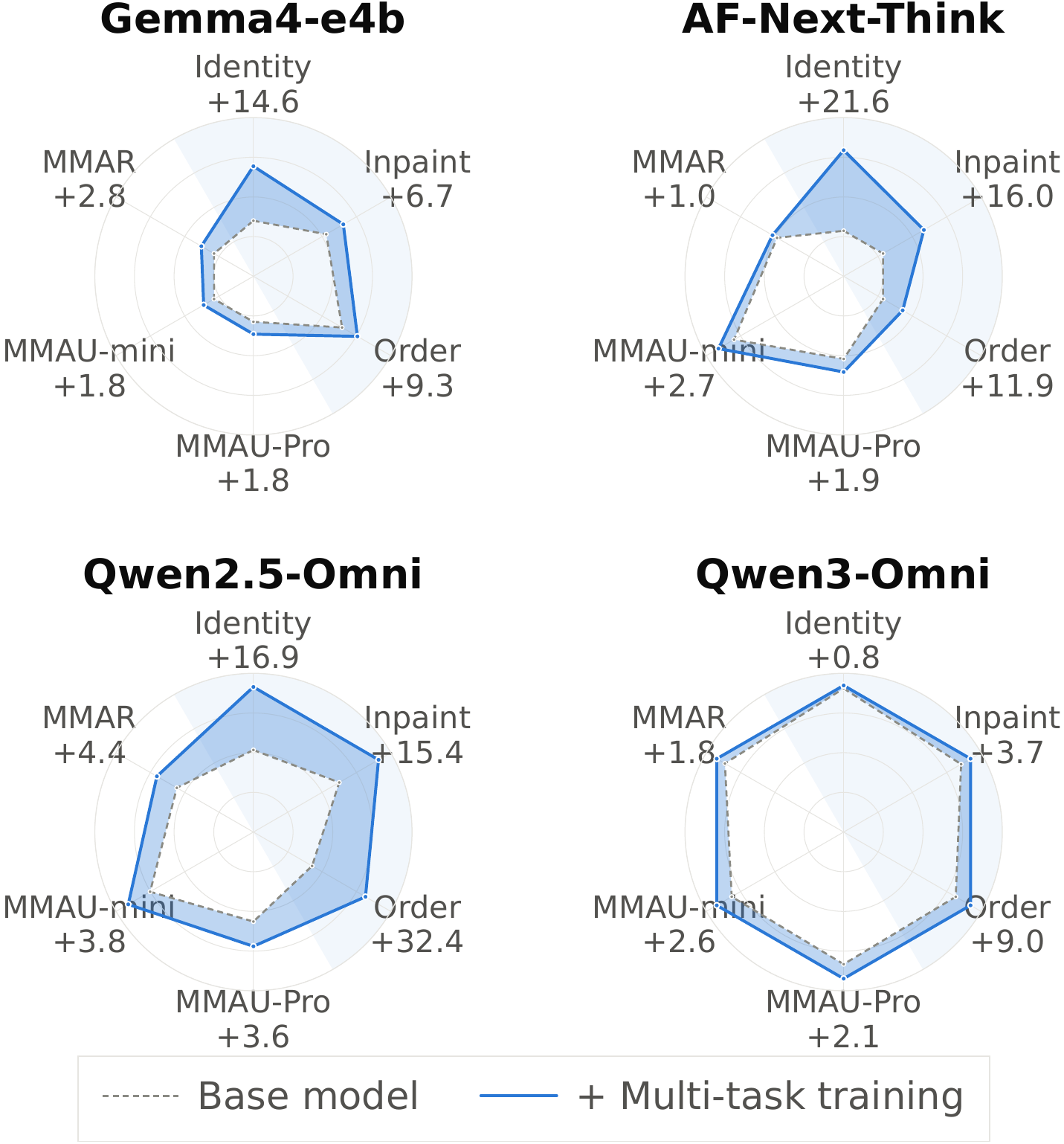}
    \caption{
    \textbf{Accuracy comparison before and after Audio-R1 across different LALMs.} Radar plots show performance on three self-supervised tasks and three downstream benchmarks. 
    }
    \label{fig:backbone_comparison}
    \vspace{-4mm}
\end{figure}

\section{Results}

We evaluate Audio-R1 by addressing three research questions that probe its effectiveness and transferability.
First, we ask whether Audio-R1 consistently improves performance across different audio-language model backbones (\textbf{Q1}).
We then investigate how individual self-supervised tasks contribute to capability acquisition and benchmark generalization, and whether combining them provides complementary benefits (\textbf{Q2}).
Finally, we compare Audio-R1 with SFT to determine how the two training paradigms differ in downstream generalization and the quality of the reasoning traces (\textbf{Q3}).

\begin{table}[t]
\centering
\caption{
\textbf{Accuracy of Qwen2.5-Omni under single-task and multi-task self-supervised training.}
We report performance on three self-supervised task capabilities and the \textit{average} across three downstream benchmarks: MMAR, MMAU-mini, and MMAU-Pro.
We report post-training results as mean $\pm$ std over three independent training runs with different random seeds.
}
\label{tab:master_transfer_matrix}
\small
\setlength{\tabcolsep}{3.8pt}
\renewcommand{\arraystretch}{1.12}

\resizebox{\columnwidth}{!}{
\begin{tabular}{lccc|c}
\toprule
\textbf{Model / Training}
&
\multicolumn{3}{c|}{\textbf{Self-supervised Capability}}
&
\textbf{Benchmarks}\\

\cmidrule(lr){2-4}
\cmidrule(lr){5-5}

&
\textbf{Identity ($\uparrow$)}
&
\textbf{Inpainting ($\uparrow$)}
&
\textbf{Ordering ($\uparrow$)}
&
\textbf{Average ($\uparrow$)}
\\

\midrule

Base model

& 80.7
& 75.0
& 27.3

& 62.2
\\

\midrule

+ Identity Only

& \underline{95.7} ${\scriptstyle\pm 0.3}$
& 84.7 ${\scriptstyle\pm 2.2}$
& 32.5 ${\scriptstyle\pm 2.1}$

& 64.4 ${\scriptstyle\pm 0.6}$
\\

+ Inpainting Only

& 87.0 ${\scriptstyle\pm 1.5}$
& \underline{94.1} ${\scriptstyle\pm 0.5}$
& 35.7 ${\scriptstyle\pm 2.2}$

& \underline{64.8} ${\scriptstyle\pm 0.5}$
\\

+ Ordering Only

& 82.3 ${\scriptstyle\pm 2.4}$
& 88.3 ${\scriptstyle\pm 1.4}$
& \underline{53.2} ${\scriptstyle\pm 1.8}$

& 64.0 ${\scriptstyle\pm 1.2}$
\\

\rowcolor{royalblue!20}
+ Multi-task (Ours)

& \textbf{97.6} ${\scriptstyle\pm 0.4}$
& \textbf{96.4} ${\scriptstyle\pm 0.3}$
& \textbf{59.7} ${\scriptstyle\pm 1.4}$

& \textbf{66.2} ${\scriptstyle\pm 0.8}$
\\

\bottomrule
\end{tabular}
}

\vspace{-2mm}
\end{table}

\subsection{Q1: Is Audio-R1 effective and generalizable across different audio-language models?}
\label{subsec:rq1}

Table~\ref{tab:main_benchmark_comparison} shows that Audio-R1 improves performance on all three downstream audio benchmarks across all four evaluated backbones.
Notably, Audio-R1-Qwen3-Omni outperforms the evaluated closed-source, open-weight, and RL-finetuned baselines, achieving the highest scores of 70.3\% on MMAR, 77.5\% on MMAU-mini, and 63.6\% on MMAU-Pro.
Figure~\ref{fig:backbone_comparison} further shows that the evaluated backbones consistently improve on the self-supervised tasks alongside their downstream benchmark gains.
Together, these results demonstrate that Audio-R1 improves both task-specific performance and broader audio understanding across multiple model families.
Sub-category gains span related abilities, including temporal reasoning and acoustic analysis, as well as untargeted areas such as music, emotion, and knowledge-based questions (Appendix~\ref{app:subtask_gains}).

\subsection{Q2: Do the self-supervised tasks provide complementary benefits for capability learning and downstream transfer?}
\label{subsec:rq2}

Table~\ref{tab:master_transfer_matrix} examines the effect of training on individual SSL tasks and their combination.
For a controlled comparison, we adjust the number of epochs in each single-task run to match the total number of training steps used in multi-task training.
Among the single-task training configurations, each achieves the highest performance on its own training task.
Beyond these task-specific gains, every single-task configuration also improves performance on both other SSL tasks relative to the base model, indicating positive cross-task transfer.
Importantly, combining all three tasks consistently outperforms single-task training, achieving gains of +16.9, +15.4, and +32.4 pp on the three self-supervised capabilities and an average gain of +4.0 pp across the three downstream benchmarks.
These results suggest that the tasks provide complementary learning signals, improving task performance and downstream generalization, while reward availability shifts over training and the joint update integrates rather than simply sums task-specific updates (Appendices~\ref{app:difficulty} and~\ref{app:task_geometry}).


\begin{figure}[t]
    \centering
    \includegraphics[width=0.95\linewidth]{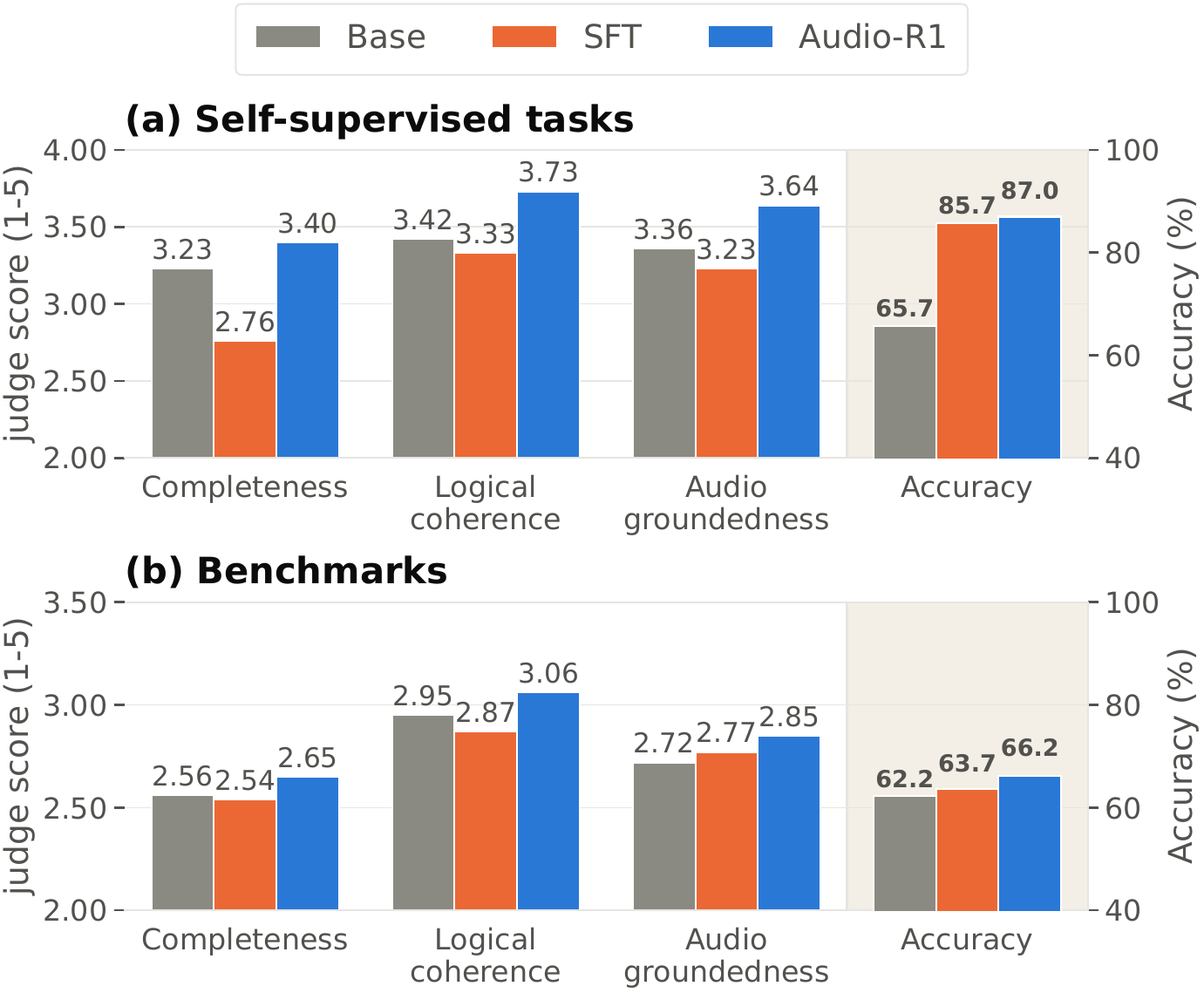}
    \caption{
    \textbf{Answer accuracy and judged reasoning quality under different learning approaches.}
    Completeness, logical coherence, and audio groundedness are scored on a 1-5 scale (left axes); answer accuracy is reported as a percentage in the shaded regions (right axes).
    }
    \label{fig:sft_vs_reasoning_scatter}
    \vspace{-4mm}
\end{figure}

\subsection{Q3: How do learning approaches differ in generalization and reasoning quality?}
\label{subsec:rq3}

To examine whether the constructed self-supervised data alone accounts for the observed gains, we train an SFT baseline using the same task instances and derived answers.
To evaluate generated reasoning, we use Gemini 3.5 Flash as an audio-conditioned LLM-as-judge, providing the audio, question and choices, reasoning trace, and reference answer, together with construction metadata for SSL tasks.
The judge scores three dimensions on a 1-5 scale: \emph{completeness}, whether the trace provides sufficient evidence and inferential steps; \emph{logical coherence}, whether its reasoning is consistent and supports its conclusion; and \emph{audio groundedness}, whether its audio-related claims are supported by the recording.

Figure~\ref{fig:sft_vs_reasoning_scatter} compares answer accuracy and generated reasoning quality for the base model, SFT, and Audio-R1.
Although both methods produce comparable gains on self-supervised tasks, SFT provides only modest downstream transfer, whereas Audio-R1 achieves substantially larger downstream improvements.
Thus, higher self-supervised task accuracy alone does not guarantee better benchmark performance.
Moreover, the quality of the reasoning scores reveal a related contrast.
Audio-R1 achieves higher mean scores than both the base model and SFT across all three dimensions in both evaluations.
SFT, in contrast, generally yields lower reasoning-quality scores than the base model, suggesting that answer-based supervision can degrade the model's ability to produce high-quality reasoning traces  (detailed in Appendix~\ref{app:sft_rl_ablation}).
These results indicate that the proposed self-supervised task data alone induces some transfer under SFT, while the RL objective in Audio-R1 amplifies this transfer, highlighting the importance of both \emph{what} capability is learned and \emph{how} it is learned.
Examples of generated reasoning are provided in Appendix~\ref{app:reasoning_examples}.

\section{Conclusion}
\label{sec:conclusion}

We presented Audio-R1, an RL post-training framework that derives verifiable rewards from three self-supervised audio tasks without human annotations or external supervision.
Experiments showed consistent gains in audio understanding across four LALM backbones, while task ablations highlighted the benefits of multi-task training with complementary learning signals.
These findings support self-supervised RL post-training as a promising direction for developing more general-purpose audio capabilities.
We encourage researchers to develop various methodologies to take advantage of grounding in audio reasoning.
Future work can explore richer self-supervised tasks, adaptive task selection and mixing, and generalization to more diverse real-world audio settings.

\vfill
\pagebreak
\section{Acknowledgments}
We acknowledge the support of the Natural Sciences and Engineering Research Council of Canada (NSERC) and the Digital Research Alliance of Canada (alliancecan.ca). Cem Subakan, and Mirco Ravanelli are respectively supported by the NSERC projects RGPIN-2023-05759, and RGPIN-2022-05088. Mirco Ravanelli also is supported by Translated through the Imminent research grant.


\bibliographystyle{IEEEbib}
\bibliography{strings,refs}

\newpage
\appendix
\onecolumn

\begin{center}
    {\Large \textbf{Audio-R1: Audio Reasoning through Reinforcement Learning on \\ Self-Supervised Tasks}}\\
    \vspace{0.5em}{\Large Appendix} \\
    \vspace{1.5em}
\end{center}

\section{Task and Judge Instructions}
\label{app:instructions}

\subsection{Task Prompts}



\begin{tcolorbox}[colback=gray!4, colframe=gray!45, arc=1.5mm, boxrule=0.4pt, left=2mm, right=2mm, top=1mm, bottom=1mm, title={\small\textbf{Identity Detection}}, coltitle=black, colbacktitle=gray!15, breakable]
\footnotesize\ttfamily\raggedright
You are given an anchor audio clip and two candidate audio clips:\\
Anchor: <audio>\\
Candidate 1: <audio>\\
Candidate 2: <audio>\\
For each candidate, decide whether it is the same recording as the Anchor (an exact copy) or comes from a different source recording.\\
Which option is correct?\\
(A) Only Candidate 1 is the same recording as the Anchor (Candidate 2 is from a different recording).\\
(B) Only Candidate 2 is the same recording as the Anchor (Candidate 1 is from a different recording).\\
(C) Both Candidate 1 and Candidate 2 are the same recording as the Anchor.\\
(D) Neither candidate is the same recording as the Anchor.\\

\end{tcolorbox}
\begin{tcolorbox}[colback=gray!4, colframe=gray!45, arc=1.5mm, boxrule=0.4pt, left=2mm, right=2mm, top=1mm, bottom=1mm, title={\small\textbf{Region Inpainting}}, coltitle=black, colbacktitle=gray!15, breakable]
\footnotesize\ttfamily\raggedright
You are given an audio clip with one missing part.\\
<audio>\\
\medskip
Task:\\
Determine the candidate audio segment that best fits the missing part of the original audio.\\
Select the correct answer from the following 4 choices:\\
(A) <audio>\\
(B) <audio>\\
(C) <audio>\\
(D) <audio>\\
\medskip
The missing part should be inferred from the surrounding audio context, such as speech content, speaker identity, acoustic continuity, background sound, and temporal flow.\\
\end{tcolorbox}
\begin{tcolorbox}[colback=gray!4, colframe=gray!45, arc=1.5mm, boxrule=0.4pt, left=2mm, right=2mm, top=1mm, bottom=1mm, title={\small\textbf{Temporal Ordering}}, coltitle=black, colbacktitle=gray!15, breakable]
\footnotesize\ttfamily\raggedright
You are given 3 audio segments cut from a single continuous conversation.\\
They are played in a randomly shuffled order and numbered by that playback order:\\
\medskip
Segment 1: <audio>\\
Segment 2: <audio>\\
Segment 3: <audio>\\
\medskip
Your task is to recover the ORIGINAL chronological order of the 3 segments. Each option below lists the playback segment indices in chronological order.\\
\medskip
Reason step by step:\\
Step 1 -- Transcribe each segment in full, labeled by its playback number.\\
Step 2 -- Find cues that link the segments: question-answer pairs, discourse markers, pronoun references, topic and turn continuity, sentences that begin in one segment and finish in another, speaker turn-taking, and prosody.\\
Step 3 -- Decide which playback segment came first, second, and so on, then match your conclusion to one of the options.\\
\medskip
Which option gives the correct chronological order?\\
(A) 2, 3, 1\\
(B) 3, 2, 1\\
(C) 1, 3, 2\\
(D) 2, 1, 3\\
\end{tcolorbox}

\subsection{Judge Prompt}

The judge receives the audio clip(s), the question and answer choices, the reference answer, and the model's reasoning chain (the text inside \texttt{<think>} and \texttt{</think>}).
For the self-supervised tasks, it additionally receives the construction metadata, such as the true segment order, the mapping from options to orders, the source identities of the clips, and the location of the masked region.
The system instruction is reproduced verbatim below.
The judge also returns \texttt{overall} score, we report only the three dimensions in Fig.~\ref{fig:sft_vs_reasoning_scatter}.

\begin{tcolorbox}[colback=gray!4, colframe=gray!45, arc=1.5mm, boxrule=0.4pt, left=2mm, right=2mm, top=1mm, bottom=1mm, title={\small\textbf{Judge system instruction}}, coltitle=black, colbacktitle=gray!15, breakable]
\footnotesize\ttfamily\raggedright
You are an expert judge of REASONING-CHAIN quality for audio question-answering.\\
\medskip
You will receive: the audio clip(s), a multiple-choice question, the model's step-by-step reasoning chain, and -- when the benchmark is synthetically constructed -- ORACLE ground-truth metadata describing exactly how the audio was built (true segment order, choice->order mapping, source/speaker identities, inpaint location, etc.).\\
\medskip
Listen to the audio and evaluate ONLY the quality of the reasoning chain. Do NOT judge whether the final answer letter is correct -- that is scored separately by a deterministic verifier, so ignore final-answer correctness and focus on the reasoning itself.\\
\medskip
Score each dimension from 1 (very poor) to 5 (excellent):\\
\medskip
- completeness: Are the evidence and intermediate steps needed to support the conclusion sufficient? All relevant segments/events/sub-questions actually addressed, with no missing links.\\
- logical\_coherence: Are the claims free of internal contradiction, and does the conclusion follow validly from the stated evidence (no non-sequiturs, and no mismatch between the reasoning and the conclusion it draws)?\\
- audio\_groundedness: Are the specific audio-related claims actually supported by what is audible in the clip, rather than generic priors, guessed content, or assertions you cannot verify from the audio? Fluent but unsupported claims score LOW even if they sound plausible.\\

\medskip
Also output:\\
- overall: a 1-5 holistic judgment of the reasoning chain's quality, independent of whether the final answer letter is correct.\\
\medskip
Respond with ONLY a JSON object, no prose, no code fences:\\
\{"completeness": int, "logical\_coherence": int, "audio\_groundedness": int, "construction\_consistency": int or null, "overall": int\}\\
\end{tcolorbox}

\section{Benchmark Gains by Subtask}
\label{app:subtask_gains}

The benchmark averages in Table~\ref{tab:main_benchmark_comparison} could hide gains concentrated in a few question types.
To check this, we break the results down by the categories that each benchmark defines.
Table~\ref{tab:app_subtask_fine} shows this breakdown for Qwen2.5-Omni, the backbone used in our ablations.

\begin{table}[H]
\centering
\caption{Change in Qwen2.5-Omni accuracy (percentage points) after Audio-R1 for every MMAU-mini sub-category with at least 20 items, every MMAR sub-category with at least 15 items, and every MMAU-Pro category with at least 40 multiple-choice items, sorted by gain.}
\label{tab:app_subtask_fine}
\footnotesize
\setlength{\tabcolsep}{4pt}
\scalebox{0.98}{%
\begin{tabular}{lrr|lrr}
\toprule
\textbf{Sub-category} & $n$ & $\Delta$ & \textbf{Sub-category} & $n$ & $\Delta$ \\
\midrule
\textbf{MMAU-mini} (73.4\,$\rightarrow$\,77.2\%) & 1000 & \textbf{+3.8} & \textbf{MMAR} (58.2\,$\rightarrow$\,62.6\%) & 994 & \textbf{+4.4} \\
Temporal Reasoning & 56 & \textbf{+18.0} & Spatial Analysis & 15 & \textbf{+10.8} \\
Phonemic Stress Pattern Analysis & 53 & \textbf{+11.7} & Acoustic Quality Analysis & 18 & \textbf{+9.0} \\
Acoustic Scene Reasoning & 48 & \textbf{+8.1} & Professional Knowledge and Reasoning & 71 & \textbf{+8.1} \\
Key highlight Extraction & 21 & \textbf{+7.4} & Emotion and Intention & 60 & \textbf{+6.8} \\
Dissonant Emotion Interpretation & 35 & \textbf{+6.7} & Correlation Analysis & 50 & \textbf{+6.5} \\
Harmony and Chord Progressions & 33 & \textbf{+4.7} & Temporal Analysis & 28 & \textbf{+5.8} \\
Musical Texture Interpretation & 34 & \textbf{+4.6} & Culture of Speaker & 49 & \textbf{+5.0} \\
Socio-cultural Interpretation & 20 & \textbf{+3.9} & Environmental Perception and Reasoning & 149 & \textbf{+4.9} \\
Emotion Flip Detection & 20 & \textbf{+3.9} & Anomaly Detection & 17 & \textbf{+4.8} \\
Rhythm and Tempo Understanding & 46 & \textbf{+3.3} & Counting and Statistics & 99 & \textbf{+4.2} \\
Event-Based Sound Reasoning & 48 & \textbf{+3.3} & Content Analysis & 302 & \textbf{+2.7} \\
Temporal Event Reasoning & 48 & \textbf{+3.3} & Music Theory & 63 & \textbf{+1.3} \\
Ambient Sound Interpretation & 48 & \textbf{+3.3} & Speaker Analysis & 48 & +0.0 \\
Acoustic Source Inference & 48 & \textbf{+3.3} & & & \\
Counting & 29 & \textbf{+2.6} & \textbf{MMAU-Pro, MCQ} (55.1\,$\rightarrow$\,58.7\%) & 4989 & \textbf{+3.6} \\
Instrumentation & 35 & \textbf{+2.2} & Sound + music & 44 & \textbf{+11.1} \\
Emotion State summarisation & 44 & \textbf{+1.8} & Multi-audio & 426 & \textbf{+7.3} \\
Eco-Acoustic Knowledge & 47 & \textbf{+1.6} & Sound & 1027 & \textbf{+6.8} \\
Melodic Structure Interpretation & 33 & +0.0 & Voice chat & 289 & \textbf{+6.7} \\
Sound-Based Event Recognition & 46 & +0.0 & Speech & 891 & \textbf{+2.8} \\
Multi Speaker Role Mapping & 27 & +0.0 & Music & 1418 & \textbf{+1.8} \\
Conversational Fact Retrieval & 22 & +0.0 & Open & 438 & \textbf{+1.1} \\
Musical Genre Reasoning & 34 & +0.0 & Sound + speech & 88 & \textbf{+1.1} \\
Emotional Tone Interpretation & 33 & +0.0 & Spatial audio & 315 & \textbf{+0.6} \\
Phonological Sequence Decoding & 49 & -1.6 & Music + speech & 46 & -2.1 \\
Event-Based Knowledge Retrieval & 33 & -2.3 & & & \\
\bottomrule
\end{tabular}
}
\end{table}

\begin{figure}[H]
\centering
\caption{Sub-category improvement counts and overall accuracy gains for the other three backbones after Audio-R1.}
\label{fig:app_subtask_breadth}
\includegraphics[width=0.8\linewidth]{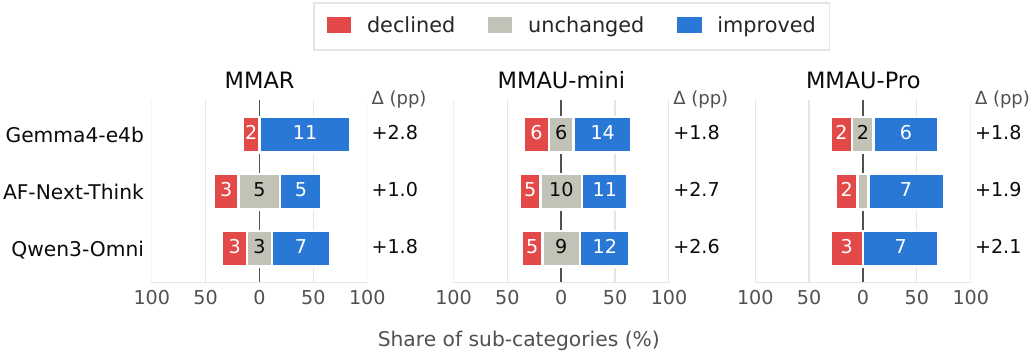}
\end{figure}

Overall, improvements are broad across MMAU-mini, MMAR, and MMAU-Pro rather than concentrated in a small set of categories (Table~\ref{tab:app_subtask_fine}).
Notably, many of the improved categories involve abilities related to those exercised by our self-supervised tasks: gains in temporal reasoning align with Temporal Ordering, while improvements in acoustic analysis are relevant to the contextual and continuity cues used in Region Inpainting.
The gains also extend to areas not directly targeted by these tasks, including music, emotion, and knowledge-based questions.
Figure~\ref{fig:app_subtask_breadth} summarizes sub-category outcomes for the other three backbones. 
Across all nine backbone--benchmark combinations, more sub-categories improve than decline, indicating a modest but consistent upward tendency. Together with the Qwen2.5-Omni results above, this pattern is observed across a broad range of audio-understanding sub-categories and backbones.

\section{Task Difficulty and Reward-Signal Density}
\label{app:difficulty}

\begin{figure}[H]
\centering
    \caption{GRPO learning signal during \textbf{multi-task training} of Qwen2.5-Omni. At each logged step, the task's prompts are split into those with mixed rewards, which receive nonzero advantages, and those whose rollouts are all wrong or all correct, which receive none.}
\label{fig:app_reward_density}
\includegraphics[width=0.8\linewidth]{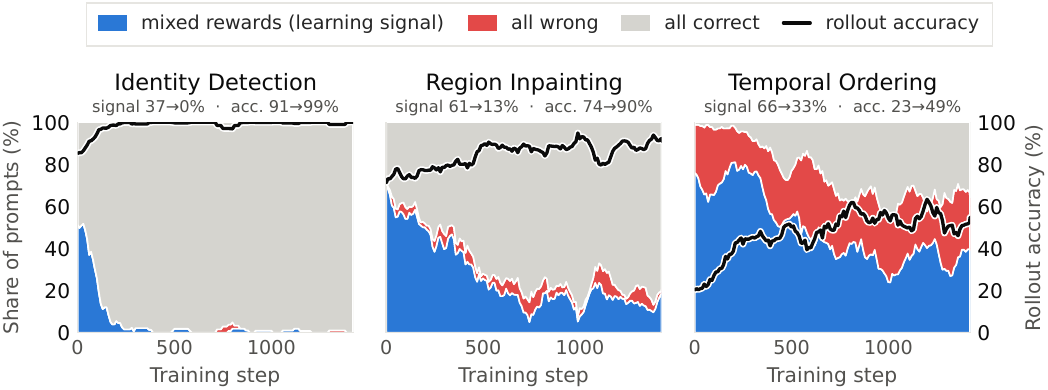}
\end{figure}

Each task presents four options, so uniform guessing is correct 25\% of the time.
For Qwen2.5-Omni, the base model is well above this level on Identity Detection (80.7\%) and Region Inpainting (75.0\%), but close to it on Temporal Ordering (27.3\%) (Table~\ref{tab:master_transfer_matrix}).
Under GRPO, a prompt contributes a learning signal only when its group of sampled responses receives different rewards.
If all responses are correct, or all are wrong, the group-normalized advantages are zero.
To measure how often this happens, we analyze the sampled completions logged during multi-task training.
Figure~\ref{fig:app_reward_density} tracks, for each task, rollout accuracy and how the proportions of prompt groups with mixed, all-wrong, or all-correct accuracy rewards change over training.

Across all three tasks, rollout accuracy trends upward over training, providing evidence that GRPO learns from the verifiable rewards. 
The trajectories nevertheless differ substantially.
Identity Detection approaches ceiling quickly, and its mixed-reward groups nearly disappear. Region Inpainting improves more gradually, with mixed-reward groups declining but persisting. 
Temporal Ordering improves from near-chance performance, but its accuracy fluctuates considerably and a substantial share of groups continues to provide mixed-reward signal late in training.
These divergent trajectories make task-specific difficulty empirically visible in this shared training run, meaning that tasks differ in learning pace and in how much challenge remains as training proceeds. 
The contrast also suggests a limitation of scaling a fixed task mixture: once a task is nearly solved, it may continue to consume rollout budget despite offering little accuracy-based learning signal. 
An adaptive curriculum could shift later-stage updates toward tasks with persistent mixed outcomes, treating reward variation as an indicator of available accuracy-based signal.

\section{Single-Task and Joint-Training Update Relationships}
\label{app:task_geometry}

\begin{figure}[H]
    \centering
    \includegraphics[width=0.38\textwidth]{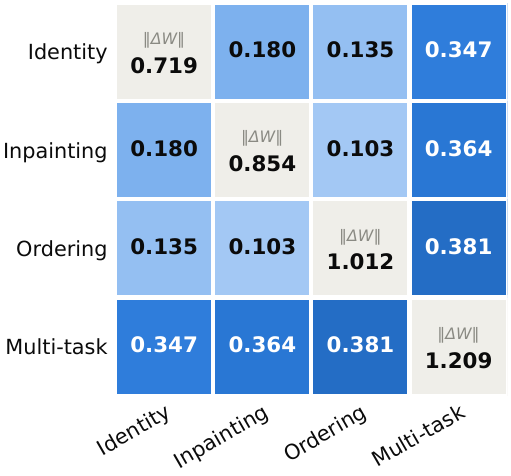}
    \caption{Pairwise cosine similarities and norms of parameter updates from single-task and multi-task training. Off-diagonal entries show cosine similarities, while diagonal entries show update norms $\|\Delta W\|$.}
    \label{fig:app_task_geometry}
\end{figure}

Whereas the reward-density analysis tracks how often sampled response groups yield nonzero GRPO advantages during training (Appendix~\ref{app:difficulty}), this analysis compares the directions and magnitudes of parameter updates from single-task runs with those from joint training.
Figure~\ref{fig:app_task_geometry} provides a parameter-space view of how joint training relates to learning on each task in isolation.
Comparing update directions and magnitudes shows whether the joint update aligns with task-specific updates or reflects a different balance among them; all updates are measured from the same base model.

Let $\theta_0$ denote the shared initialization, $\theta_i$ the parameters after single-task training on task $i$, and $\theta_{\mathrm{joint}}$ those after joint training; define $\Delta_i=\theta_i-\theta_0$ and $\Delta_{\mathrm{joint}}=\theta_{\mathrm{joint}}-\theta_0$.
We compare the pairwise cosine similarities $\cos(\Delta_i,\Delta_j)$ among single-task updates with $\cos(\Delta_{\mathrm{joint}},\Delta_i)$ between the joint and each single-task update, as well as their norms.
The single-task updates are only weakly aligned with one another, whereas the joint update is more closely aligned with each of them.
Its norm satisfies $\max_i \|\Delta_i\|_2 < \|\Delta_{\mathrm{joint}}\|_2 < \|\sum_i \Delta_i\|_2$, indicating that joint training combines the task directions with a changed balance rather than simply adding independently learned updates.
Together, these diagnostics provide complementary evidence on the integration of task-specific learning signals in multi-task RL.

\section{SFT and RL: Accuracy and Reasoning Quality}
\label{app:sft_rl_ablation}

The SFT baseline is trained on the same self-supervised instances and construction-derived answers as Audio-R1.
The two runs therefore differ only in the learning objective.
The Audio-R1 task-level accuracy entries match the three-run means in Table~\ref{tab:master_transfer_matrix}, while the judge scores correspond to Fig.~\ref{fig:sft_vs_reasoning_scatter}.
For the self-supervised tasks, accuracy and judge scores cover every test item.
For the benchmarks, accuracy is the mean over the full MMAR, MMAU-mini, and MMAU-Pro test sets.

\begin{table}[h]
\centering
\caption{Accuracy (\%) and judge scores (1--5) for Qwen2.5-Omni before training (Base), after SFT, and after Audio-R1, all trained on the same self-supervised data. (C = completeness, L = logical coherence, G = audio groundedness)}
\label{tab:app_sft_rl}
\small
\setlength{\tabcolsep}{4pt}
\begin{tabular}{l ccc ccc ccc ccc}
\toprule
\textbf{Evaluation} & \multicolumn{3}{c}{\textbf{Accuracy}} & \multicolumn{3}{c}{\textbf{C}} & \multicolumn{3}{c}{\textbf{L}} & \multicolumn{3}{c}{\textbf{G}} \\
\cmidrule(lr){2-4}\cmidrule(lr){5-7}\cmidrule(lr){8-10}\cmidrule(lr){11-13}
& Base & SFT & Ours & Base & SFT & Ours & Base & SFT & Ours & Base & SFT & Ours \\
\midrule
Identity   & 80.7 & 94.1 & \textbf{97.6} & 3.89 & 3.60 & \textbf{4.20} & 4.15 & 4.23 & \textbf{4.48} & 3.86 & 3.97 & \textbf{4.25} \\
Inpainting & 75.0 & 86.1 & \textbf{96.4} & 3.35 & 2.31 & \textbf{3.45} & 3.78 & 3.10 & \textbf{4.08} & 3.57 & 3.08 & \textbf{3.81} \\
Ordering   & 27.3 & \textbf{71.4} & 59.7 & \textbf{2.05} & 1.79 & \textbf{2.05} & 1.89 & 2.05 & \textbf{2.18} & 2.36 & 2.16 & \textbf{2.48} \\
\midrule
Self-supervised Tasks & 65.7 & 85.7 & \textbf{87.0} & 3.23 & 2.76 & \textbf{3.40} & 3.42 & 3.33 & \textbf{3.73} & 3.36 & 3.23 & \textbf{3.64} \\
Benchmarks & 62.2 & 63.7 & \textbf{66.2} & 2.56 & 2.54 & \textbf{2.65} & 2.95 & 2.87 & \textbf{3.06} & 2.72 & 2.77 & \textbf{2.85} \\
\bottomrule
\end{tabular}
\end{table}

The key distinction is not whether SFT can raise answer accuracy but whether that accuracy comes with a well-supported reasoning chain.
Despite its strong task accuracy, SFT produces substantially shorter traces and receives lower completeness scores on all three tasks, as well as lower audio-groundedness scores on Inpainting and Ordering.
This pattern suggests that learning from answer-targeted traces may improve answer production without preserving the ability to generate complete, audio-grounded reasoning chains; it is suggestive evidence, not proof of a causal loss in reasoning ability.
Audio-R1's advantage is that it combines strong overall accuracy and greater benchmark transfer (+4.0 vs. +1.5 points) with higher logical-coherence scores than SFT on all three tasks and stronger scores on the other judge dimensions.

\section{Reasoning Examples}
\label{app:reasoning_examples}

We reproduce verbatim outputs before training, after SFT, and after Audio-R1 on held-out self-supervised test items.
Each title gives the model's answer and its judge scores (C/L/G as in Table~\ref{tab:app_sft_rl}).
Example~1 compares Base and Audio-R1, and Example~2 adds SFT.
Example~3 is a failure case of Audio-R1.

\medskip
\noindent\textbf{Example 1: Identity Detection (Audio-R1 corrects the base model).}
Options: (A) only Candidate 2 matches the anchor, (B) only Candidate 1 matches, (C) neither matches, (D) both match. Reference answer: (C).
The base model treats a gunshot-like sound effect, related to the anchor's topic, as evidence of the same recording. Audio-R1 instead rejects both candidates because neither matches the anchor's content.

\begin{tcolorbox}[colback=white, colframe=gray!45, arc=1.5mm, boxrule=0.4pt, left=2mm, right=2mm, top=1mm, bottom=1mm, title={\small Base --- answer A (\textcolor{red!70!black}{incorrect}); judge C/L/G = 1/1/1}, coltitle=black, colbacktitle=gray!10, breakable]
\footnotesize\raggedright
\texttt{<think>}\\
To determine which candidate is the same recording as the anchor, we need to compare the content and context of the audio clips. The anchor audio clip discusses the assassination of Robert F. Kennedy and its impact. \\
\smallskip
- Candidate 1 contains a speech with the phrase ``Well, despite the yummy bagel,'' which is unrelated to the content of the anchor clip.\\
- Candidate 2 contains a sound effect that resembles a gunshot, which could be related to the assassination mentioned in the anchor clip.\\
\smallskip
Given this analysis, Candidate 2 seems to be the same recording as the anchor, as it contains a sound effect that could be related to the assassination mentioned in the anchor clip.\\
\texttt{</think>}\\
\texttt{<answer>}A\texttt{</answer>}\\
\end{tcolorbox}
\begin{tcolorbox}[colback=white, colframe=gray!45, arc=1.5mm, boxrule=0.4pt, left=2mm, right=2mm, top=1mm, bottom=1mm, title={\small Audio-R1 --- answer C (\textcolor{green!45!black}{correct}); judge C/L/G = 5/5/5}, coltitle=black, colbacktitle=gray!10, breakable]
\footnotesize\raggedright
\texttt{<think>}\\
To determine which candidate is the same recording as the anchor, we need to compare the content and context of the audio clips. The anchor audio clip discusses the assassination of Robert F. Kennedy and its impact. \\
\smallskip
- Candidate 1 contains a speech about yummy bagels, which is unrelated to the content of the anchor.\\
- Candidate 2 contains a sound effect that is not related to the content of the anchor.\\
\smallskip
Neither candidate matches the content of the anchor, so neither is the same recording as the anchor.\\
\texttt{</think>}\\
\texttt{<answer>}C\texttt{</answer>}\\
\end{tcolorbox}

\medskip
\noindent\textbf{Example 2: Region Inpainting (SFT gives no answer).}
Four candidate segments; reference answer: (B).
The base model and SFT both conclude that no candidate fits; SFT leaves the answer tag empty.
According to its trace, Audio-R1 uses the spoken context (``the signing of the Treaty of'' \ldots\ ``World War'') to select the candidate containing ``Versailles''.

\begin{tcolorbox}[colback=white, colframe=gray!45, arc=1.5mm, boxrule=0.4pt, left=2mm, right=2mm, top=1mm, bottom=1mm, title={\small Base --- answer A (\textcolor{red!70!black}{incorrect}); judge C/L/G = 1/2/2}, coltitle=black, colbacktitle=gray!10, breakable]
\footnotesize\raggedright
\texttt{<think>}\\
To determine the correct answer, I need to analyze the context of the given audio clip and the provided options. The audio clip mentions ``the signing of the Treaty of'' and ``World War,'' which suggests a historical context, likely referring to a significant event in history. The options provided do not seem to fit the historical context of the audio clip. Therefore, none of the options appear to be the correct answer based on the given information.\\
\texttt{</think>}\\
\texttt{<answer>}A\texttt{</answer>}\\
\end{tcolorbox}
\begin{tcolorbox}[colback=white, colframe=gray!45, arc=1.5mm, boxrule=0.4pt, left=2mm, right=2mm, top=1mm, bottom=1mm, title={\small SFT --- answer none (\textcolor{red!70!black}{incorrect}); judge C/L/G = 1/1/1}, coltitle=black, colbacktitle=gray!10, breakable]
\footnotesize\raggedright
\texttt{<think>}\\
The audio clip provided is a speech segment that mentions the signing of a treaty, which is a significant historical event. The options given are unrelated to the context of the speech. Therefore, the correct answer is not among the provided choices.\\
\texttt{</think>}\\
\texttt{<answer>}\texttt{</answer>}\\
\end{tcolorbox}
\begin{tcolorbox}[colback=white, colframe=gray!45, arc=1.5mm, boxrule=0.4pt, left=2mm, right=2mm, top=1mm, bottom=1mm, title={\small Audio-R1 --- answer B (\textcolor{green!45!black}{correct}); judge C/L/G = 4/4/5}, coltitle=black, colbacktitle=gray!10, breakable]
\footnotesize\raggedright
\texttt{<think>}\\
The task requires identifying the most suitable audio segment to fill in the missing part of the original audio. The context provided is about the signing of a treaty, which is a significant historical event. The options given are unrelated to this context, so we need to focus on the historical event mentioned in the original audio clip. The correct answer should be related to the Treaty of Versailles, which ended World War I. Option B mentions ``Versailles,'' which is the correct choice.\\
\texttt{</think>}\\
\texttt{<answer>}B\texttt{</answer>}\\
\end{tcolorbox}

\medskip
\noindent\textbf{Example 3: Temporal Ordering (failure case).}
Options: (A) 3, 1, 2, (B) 2, 1, 3, (C) 3, 2, 1, (D) 2, 3, 1. Reference answer: (B).
The base model transcribes all three segments and answers correctly.
Audio-R1 gives a shorter trace whose stated order (2, 1, 3) is correct, but it then outputs (D), which is the order 2, 3, 1.
This mismatch between the reasoning and the final answer is the kind of error that outcome-only rewards do not penalize directly (Appendix~\ref{app:limitations}).

\begin{tcolorbox}[colback=white, colframe=gray!45, arc=1.5mm, boxrule=0.4pt, left=2mm, right=2mm, top=1mm, bottom=1mm, title={\small Base --- answer B (\textcolor{green!45!black}{correct}); judge C/L/G = 5/5/5}, coltitle=black, colbacktitle=gray!10, breakable]
\footnotesize\raggedright
\texttt{<think>}\\
To determine the correct chronological order, let's analyze the segments:\\
\smallskip
1. Segment 1: ``Do you remember the steps? Yes, I believe so.''\\
   - This segment seems to be a response to a question about remembering steps, suggesting it might be part of a conversation where someone is being asked about their memory of certain procedures.\\
\smallskip
2. Segment 2: ``Alright, today we'll focus on takeoff procedures.''\\
   - This segment appears to be an introductory statement, possibly setting the context for the discussion about takeoff procedures.\\
\smallskip
3. Segment 3: ``First we complete the preflight checklist, right?''\\
   - This segment seems to be a continuation of the discussion, specifically addressing the preflight checklist, which is a part of the takeoff procedures.\\
\smallskip
Given the context and the flow of the conversation, it is logical that the conversation starts with an introduction to the topic (Segment 2), followed by a question about remembering steps (Segment 1), and then a discussion about the preflight checklist (Segment 3).\\
\smallskip
Therefore, the correct chronological order is:\\
2, 1, 3\\
\texttt{</think>}\\
\texttt{<answer>}B\texttt{</answer>}\\
\end{tcolorbox}
\begin{tcolorbox}[colback=white, colframe=gray!45, arc=1.5mm, boxrule=0.4pt, left=2mm, right=2mm, top=1mm, bottom=1mm, title={\small Audio-R1 --- answer D (\textcolor{red!70!black}{incorrect}); judge C/L/G = 2/3/3}, coltitle=black, colbacktitle=gray!10, breakable]
\footnotesize\raggedright
\texttt{<think>}\\
Segment 1 seems to be a response to a question, which suggests it's the second segment in the conversation. Segment 2 appears to be an introductory statement, likely the first segment. Segment 3 seems to be a continuation of the conversation, possibly the third segment. Therefore, the correct order should be 2, 1, 3.\\
\texttt{</think>}\\
\texttt{<answer>}D\texttt{</answer>}\\
\end{tcolorbox}

\section{Limitations}
\label{app:limitations}

\textbf{Task coverage.} Audio-R1 uses three automatically constructed, candidate-based tasks that provide scalable labels but cover only selected forms of audio correspondence, contextual completion, and temporal ordering. Exact-match and other construction-specific cues may not be available in open-ended audio understanding, so transfer to established benchmarks does not by itself establish robustness to broader real-world distributions.

\noindent
\textbf{Reasoning evaluation.} The reward verifies final-answer correctness and format, not intermediate reasoning; a correct answer can therefore accompany an incomplete or unfaithful trace. The LLM judge provides complementary evidence, but remains a proxy and cannot establish trace faithfulness or a causal role for reasoning in accuracy gains.

\noindent
\textbf{Post-training RL limitations.} Because Audio-R1 is a post-training RL method, its gains remain conditioned on the pretrained backbone and its prior capabilities. Moreover, our experiments deliberately select tasks for which the base models already possess usable capabilities; they therefore do not establish that RL can develop reasoning skills absent from pretraining or disentangle newly acquired reasoning from inherited capabilities.

\end{document}